\documentclass[11pt,twoside,letterpaper]{article} 
\usepackage{natbib}
\usepackage{times,fancyhdr}
\usepackage[dvips]{graphicx}
\usepackage{amsmath, amsthm, amssymb}
\usepackage[latin1]{inputenc}
\usepackage{subfigure}
  
\makeatletter
\def\cleardoublepage{\clearpage\if@twoside \ifodd\c@page\else%
\hbox{}%
\thispagestyle{empty}%
\newpage%
\if@twocolumn\hbox{}\newpage\fi\fi\fi}
\makeatother

\def\figurename{Figure}
\makeatletter
\renewcommand{\fnum@figure}[1]{\figurename~\thefigure.}
\makeatother
\def\tablename{Table}
\makeatletter
\renewcommand{\fnum@table}[1]{\tablename~\thetable.}
\makeatother

\begin{document}

\title{Positive steady state in multi-species models of real food webs}
\author{Alexander Pimenov \thanks{Corresponding author: pimenov\_a@mail.ru}\\
Weierstrass Institute, Mohrenstrasse 39, D-10117 Berlin, Germany}
\maketitle
\begin{abstract}
Real food web data available in the literature presents us with the relations between various species, sizes of these species, metabolic types of the species and other useful information, which allows us to define parameters for the mathematical dynamical models of these food webs, and perform theoretical studies of these models. Unfortunately, the researches often face the problem of the extinction of the species in such situations, which could be an important limiting factor. In this paper, we propose a simple algorithm of parameterisation that leads to the existence of positive steady state and improves persistence of the species in multi-species models of real food webs.
\end{abstract}
\section{Introduction}

Mathematical modelling of interactions between species in ecological communities has nowadays moved far beyond simple cases of two or three interacting species, where general mathematical analysis could be performed for very wide feasible parameter ranges. For realistic food webs with large number of species rigorous mathematical methods are generally limited to very basic considerations such as topological analysis. Therefore, direct numerical simulations and stability analysis of the corresponding dynamical models can provide much more detailed information on the possible dynamics of the system. In particular, by picking randomly parameters from particular ranges according to Monte-Carlo procedure, we can gain some general insights in the behaviour of the system.  For that, it is extremely important to define realistic parameter ranges and initial conditions for large systems of differential equations with hundreds of parameters, which we use as our mathematical models.
The most general and practical way to do this is using statistical data for real food webs. Numerous efforts of scientists in gathering field data have lead to the derivation of statistical relationships between traits of different species, with metabolic group (vertebrate/invertebrate, cold and warm-blooded etc) and size being the most important ones, and properties of the individuals and the populations such as birth/death rates, metabolic rates, abundance (\cite{peters1986ecological,brown2004toward, mccoy2008predicting}). These relationships provide a general and robust way for defining statistically significant parameter ranges for mathematical models of species interactions, and are used in numerous theoretical studies of model and real food webs (\cite{yodzis1992body, brose2006allometric, brose2006consumer,berg2015ecological,schneider2016animal,allhoff2014evolutionary}). Unfortunately, it was found that random parameterisations of predator-prey models of real food webs with Holling type I (generalised Lotka-Volterra, or GLV) and type II functional responses lead to the lack of persistence of the species so that some of them die out during the simulations (\cite{brose2006allometric}). This problem is perceived by many researches as a natural consequence of the fact that these predator-prey models neglect a lot of important effects such as openness and spatial inhomogeneity of the habitat. However, there is very little understanding of the exact biological or mathematical reasons behind the observed instability of real food-web models.  

In this paper, we demonstrate that the absence of positive equilibrium (steady state) in such GLV models is an important factor, which leads automatically to the extinction of the species in real food webs, because it breaks a necessary condition for the persistence of the species in the system. We discuss some basic properties of the model that need to be accounted for in order to gain feasible parameterisation. We show that it is possible to parameterise GLV models for the food webs available in the literature at a positive equilibrium and gain persistence using a very simple method. 
Finally, we observe lack of persistence of the predator-prey systems with Holling type II functional response, which is prone to the paradox of enrichment, and the stabilising role of Holling type III functional response, which helps to prevent prey species from extinction at low densities. This result is in agreement with previous research (\cite{brose2006allometric, kalinkat2013body}), and, as it was recently shown (\cite{kalinkat2013body, schneider2016animal}), the role of the Hill coefficient is quite complicated in realistic food webs and must be properly addressed. Indeed, the functional response in GLV models is linear (per capita) and it reflects only the sign of the relationship (positive or negative) for each species and the strength of this relationship without any particular assumption about the particular behaviour of the species, and can be derived from a great variety of simplifying initial assumptions. On the contrary, the models with more complicated functional responses rely on the particular restrictive assumptions (such as sensory Holling disk and empirically measured functional responses), and basic intuition suggests that they generally should be used for real food webs together with advanced understanding of the behaviour for all the species that belong to the food web (\cite{kalinkat2013body,allhoff2014evolutionary,schneider2016animal}). Since any kind of model provides only a useful caricature of the dynamics of real complex food webs, the GLV model remains competitive in providing useful insights into general properties of the food webs (\cite{berg2015ecological,jonsson2015reliability,jonsson2015context}), and our approach facilitates the application of the GLV model for the study of real food webs.

\section{Metabolic theory and dynamics of food webs}
\subsection{Single consumer species} \label{sec:single}
Behaviour of an individual consumer can be effectively described using energy processor analogy (\cite{yodzis1992body}): it gains energy $E_+$ by feeding from a resource (or many resources) and spends energy $E_-$ on metabolic needs (respiration etc), activity, etc. If the current balance of energy intake $\delta E =E_+ - E_- > 0$ is positive, then it can be used for growth and breeding, and if it is negative $\delta E < 0$, then the individual organism starves and looses biomass.  On the level of consumer population biomass $C$ with numerical abundance $A$, the change of biomass is expressed using the following equation
$$
\delta C = A \delta E - L A = C (f - d),
$$
where $L$ represent loss of individuals in the population due to other factors, $f$ and $d$ are mass-specific feeding and death rates. In the absence of predation and competition the role of $L$ can be neglected, and the loss of biomass can be associated with the basal energy expenditure $E_-$, which is defined by basal metabolic rate BMR and can be drawn from available statistical data (\cite{brown2004toward}) using the allometric relationship
$$
E_- \sim s^{0.75}, \quad \mbox{ or } d \approx J_{\min}= E_-/s \sim s^{-0.25}.
$$
Maximal metabolic rate MMR follows similar allometric relationship and is several times higher than BMR. Therefore, it is natural to assume that the feeding rate is a fraction of the maximal mass-specific ingestion rate $J_{\max} \sim s^{-0.25}$, which depends on the availability of the resources, and could also depend on the abundance of consumer
$$
f = \gamma J_{\max},
$$
where for Holling Type II ($n=1$) and III ($n=2$) functional responses the fraction $\gamma$ takes the form (\cite{yodzis1992body}):
$$
\gamma = \frac{R^n}{R^n+R^n_0}.
$$
Here $R$ is the abundance of the resource and $R_0$ is half-saturation density. For Holling type I functional response (GLV model) this fraction can be written as
$$
\gamma = \frac{ \max(R, 2 R_0)}{2 R_0}.
$$
Finally, assuming $\dot C = \delta C$, one can write out a differential equation for the consumer. Moreover, one can write out an equation for the resource, and obtain a closed predator-prey system for two species, $C$ and $R$ (\cite{yodzis1992body}) (see the next subsection for another example). This approach can be easily extended for food webs with many interconnected species (\cite{brose2006allometric}), however unlike for the case with two species, it is much harder to obtain persistence for all the species in real food web models.
\subsection{Infeasible parameterisation: an example} \label{sec:infeasible}
As a simplest example, we consider a Lotka-Volterra system with two consumers and resources parameterised using the general approach described in the previous subsection (\cite{yodzis1992body})
\begin{gather*}
\dot C_1 = - J_{\min, 1} C_1 + \frac{J_{\max, 1}}{2 R_{0,1}} R_1 C_1,\\
\dot C_2 = - J_{\min, 2} C_2 + \frac{J_{\max, 2}}{2 R_{0,2}} (R_1+R_2) C_2,\\
\dot R_1 = b (1 - R_1/K_1) R_1 - R_1 (\frac{J_{\max, 1} C_1}{2 R_{0,1} e_{11}} ),\\
\dot R_2 = b (1 - R_2/K_2) R_2 - R_2 (\frac{J_{\max, 1} C_1}{2 R_{0,1} e_{21}} + \frac{J_{\max, 2} C_2}{2 R_{0,2} e_{22}}).
\end{gather*}
At the steady state $\dot C_1 = 0, \dot C_2 = 0$ we obtain the following formulas for the equilibrium resource abundances
\begin{gather*}
R_1 = 2 R_{0,1} \frac{J_{\min,1}}{J_{\max,1}},\\
R_2 = 2 R_{0,2} \frac{J_{\min,2}}{J_{\max,2}} - R_1.
\end{gather*}
We can see that $R_1>0$ for any choice of positive parameters, however one can see that $R_2=0$ automatically for the most obvious choice of fixed half-saturation densities $R_{0,2}=R_{0,1}=const$ and proportions between minimum and maximum metabolic rates $\frac{J_{\min,2}}{J_{\max,2}}=\frac{J_{\min,1}}{J_{\max,1}} = const$ for similar consumer species (even with different sizes $s_1, s_2$ and actual values of death/feeding rates $J_{\min,i}, J_{\max,i}$). If $R_{0,1},R_{0,2}$ and the mentioned proportions are not exactly fixed, then, still, with random choice of parameters some of realisations will give us $R_2<0$.

According to the general theory of GLV model, existence of a positive steady state ($C_1>0,C_2>0,R_1>0,R_2>0$) is a necessary condition for the persistence of the species in any GLV system. Therefore, we see that even for this simple system Monte-Carlo method will produce some infeasible parameterisations.
Imagine now, we add more equations of the form
$$
\dot C_k = - J_{\min, k} C_k + \frac{J_{\max, k}}{2 R_{0,k}} (R_{k-1}+R_k) C_k
$$
for $k \geq 3$, where equations for $R_{k-1}, R_k$ should be updated accordingly.

We see that at the steady state $\dot C_k = 0$ the value of the resource $R_k$ is
$R_k = 2 R_{0,k} \frac{J_{\min,k}}{J_{\max,k}} - R_{k-1}$. For $k = 3$, for example, it implies that $R_1$ should be small enough so that $R_2$ is positive, however now also $R_2$ should be small enough so that $R_3 > 0$. Therefore, for large enough $k$ it would be quite hard to choose randomly the values of all the parameters from a statistical distribution so that all $R_k > 0$ at the steady state, because of the restrictions that are imposed by the structure of our predator-prey system. For many real food webs we observe very similar situation.

This simple example shows that the structure of the food web can have great influence on the relations between various parameters of the system. On the one hand, this influence can be a natural consequence of complexity, and on the other, it can make system too sensitive to the change of parameters, which probably also means that such a food web cannot be persistent under realistic assumptions (or, maybe, the sensitivity analysis should be also done with care, because these relations between parameters should be preserved by biological reasoning - for example, basal and maximal metabolic rates should be increased or decreased simultaneously). Therefore, for each experiment the researcher should perform all the necessary thorough sensitivity tests, and only then decide if he can use the parameterisation.

In this paper, our main aim is to demonstrate that any GLV model can have feasible parameterisation. For that, we note that in the above example any random and fixed  maximal ingestion rates $J_{\max}$ and half-saturation densities $R_0$ can lead to infeasible parameterisation, so we drop these notions out of consideration and try to find actual feeding rates at the positive steady state instead using different arguments. 
Only then we parameterise out-of-equilibrium dynamical models with concrete functional responses.
\subsection{Feasible parameterisation at the steady state} \label{sec:feasible}
We consider a system of $N$ species with numerical abundances $A_i$ and size $s_i$, $i=1..N$. For clarity, we assume no predation loops and weak competition for non-basal species, and if necessary these assumptions can be relaxed.

Similarly to subsection \ref{sec:single}, we consider the evolution of biomass abundance $B_i = A_i s_i$ for top-level predators
$$
\delta B_i = B_i (f_i - d_i),
$$
where $d_i=J_{\min,i}$ are death rates parameterised by basal metabolic rates, and at the steady state $\delta B_i = 0$ we have an equality between the death rates $d_i$ and the feeding rates $f_i$
$$
f_i = d_i.
$$
Feeding effort of the consumer $i$ should be distributed between its resources $j$ according to the preference $0 \leq p_{ij} \leq 1$ such that
$$
\sum_j p_{ij} = 1,
$$
so that the predator $i$ consumes prey $j$ with the total rate
$$
f_{ij} = f_i p_{ij}.
$$
In particular, $p_{ij}$ can be chosen with respect to biomass $B_j$ or numerical $B_j/s_j$ abundance of the prey as
\begin{equation}\label{eq:pref}
p_{ij} = \frac{\sigma_{ij} B_j^n}{\sum_{j} \sigma_{ij} B_j^n},
\end{equation}
where $\sigma_{ij} = 1$ or $\sigma_{ij} = 1/s_j^n$, correspondingly, if $j$ is the prey of $i$ and $\sigma_{ij}=0$, otherwise, and $n \geq 1$.
Therefore, the population level death rates $d_i$ for any other species $i$ at the next trophic levels with predators $j$ are
\begin{equation}\label{eq:Di}
d_i B_i = J_{\min,i} B_i + \sum_j c_{ij} B_i B_j  + \sum_j \frac{p_{ji}}{e_{ji}} f_j B_j,
\end{equation}
where $c_{ij}$ are intra- and interspecific competition rates and $e_{ji}$ is the efficiency of predator $j$ over prey $i$. At the steady state, feeding efforts of the species $f_i  B_i = d_i B_i$  are distributed among their prey at the next trophic level, and so on until the basal trophic level. For the basal species, we define at the steady state the birth rate as $b_i = d_i - J_{\min,i}$.
In what follows, we assume that we know the equilibrium biomass abundances $B_i$. For example, abundances $A_i$ for general class of species obey allometric relationship (\cite{peters1986ecological})
\begin{equation}\label{eq:ab}
A_i \sim s_i^{-1}.
\end{equation}
 Therefore, the equilibrium condition between feeding and death rates $f_i = d_i$ together with \eqref{eq:Di} allows us to find actual feeding rates $f_i$ for all the species, and birth rates $b_i =f_i= d_i$ for basal species.  
 
 We note that the presented approach does not handle automatically any conditions on the maximal feeding rate $J_{\max, i}$, and, in particular, $f_i/J_{\min, i}$ can be quite big (larger than 10). In this paper, we do not focus on this particular feature, and, generally speaking, it is not possible to control precisely that all the necessary parameters satisfy some empirical conditions simultaneously for complex food webs. If the proportion between maximal and minimal metabolic rates should be controlled by logical reasoning, the possible improvement of the algorithm will involve the increase of the parameter $J_{\min,i}$ closer to $f_i$ at each step of the algorithm, and recalculation of $f_i=d_i$ using \eqref{eq:Di}. In this way, we impose additional food-web-specific conditions on the space of possible parameterisations for basal metabolic rates, which is originally based on general allometric relationships.  

\subsection{Parameterisation of dynamical models}
GLV model has the form
\begin{equation} \label{GLV}
\dot B_i = r_i B_i + \sum_j a_{ij} B_i B_j,
\end{equation}
where $r_i > 0$ for basal and $r_i < 0$ for non-basal species. Let $B_i^*>0$ be the equilibrium biomass abundances, calculated in the previous subsection, and $r_i =-J_{\min,i}$ for non-basal species, $r_i = b_i$ for basal species. For competing species $a_{ij}=-c_{ij}$, and for prey $i$, predators $j$ we have
$$
a_{ij} = -\frac{p_{ji}}{e_{ji}} \frac{f_j}{B_i^*}, \quad a_{ji} =  \frac{f_j p_{ji}}{B_i^*},
$$
where $f_j$ satisfy \eqref{eq:Di} and $p_{ji}$ satisfy \eqref{eq:pref} at equilibrium. By definition, this parameterisation satisfies
$$
r_i^* B_i^* + \sum_j a_{ij} B_i^* B_j^* = 0 = \dot B_i^*.
$$
Therefore, using the values $f_i$ calculated for the steady state $B_i^*$ we obtain parameterisation of an arbitrary GLV model. 

Finally, we consider Holling-type predator-prey model of the following form
\begin{equation}\label{Holling}
\dot B_i = r_i B_i - \sum_j c_{ij} B_i B_j + \sum_j \frac{\omega_{ij} B_i B_j^n}{g_i(B_{0i},B_1,...,B_N)}-\sum_j  \frac{ \omega_{ji} B_i^n B_j}{g_j(B_{0j}, B_1,...,B_N) e_{ji}}
\end{equation}
where $g_j > 0$ for $B_{0j}, B_1, ..., B_N>0$ such that the equation $g_0=g_j(B_{0j}, B_1,...,B_N)$ has a solution $ B_{0j} = \beta(g_0, B_1,...,B_N)$ for all $g_0, B_0, B_1, ..., B_N>0$,  $n\geq 1$ and for prey $i$, predator $j$ we have $\omega_{ji}>0$. By assuming the preference $p_{ji}$ due to \eqref{eq:pref} with the same parameter $n$ and $\sigma_{ji} = \omega_{ji}$, we obtain at the steady state the following value of half-saturation density:
$$
B_{0j}=\beta(\frac{\sum_{i} \omega_{ji} (B_i^*)^n}{f_j}, B_1^*,...,B_N^*).
$$
In particular, for $g_j(B_{0j}) = B_{0j}$, $\omega_{ji} = 1$, $n=1$, $J_{\max,j}=1$ we have a GLV model \eqref{GLV}, and since $\beta(g_0)=g_0$, we obtain $a_{ji} = \frac{f_j}{\sum_{i} B_i^*}$, which is the same parameterisation as in the beginning of the subsection.
For standard Holling type II and III functional response we have $g_j = \frac{1}{J_{\max,j}}(B_{0j}^n + \sum_k \omega_{jk} B_k^n)$, hence we obtain
$$
B_{0j}^n =\sum_{k} \omega_{jk} (B_k^*)^n \frac{J_{\max,j}-f_j}{f_j}.
$$
We note that the parameters should satisfy condition $f_j < J_{\max,j}$.

These models were parameterised at the steady state with the specific assumption about the feeding preference of the predator over its prey. In fact, this assumption holds not only at the steady state but for arbitrary moment of time in the general model \eqref{Holling}:
$$
f_{ji}(t) = f_j(t) p_{ji}(t), \quad p_{ji}(t) = \frac{\omega_{ji} B_j^n(t)}{\sum_k \omega_{jk} B_k^n(t)}, \quad f_j(t) = \frac{1}{g(B_{0j}, B_1(t), .., B_N(t)}).
$$
\subsection{Improving infeasible parameterisations}
In this section, we consider GLV model \eqref{GLV} without interspecific competition or predation loops. Generally speaking, it is not easy to find a way to change some parameters in a minimal way, especially in a non-equilibrium parameterisation (e.g., see subsection \ref{sec:single}), in order to obtain feasible equilibrium, and it always makes sense to use the main method presented in this paper (see subsection \ref{sec:feasible}) from scratch instead of fixing broken parameterisation. Nevertheless, it could be useful to consider a particular situation, which demonstrates possible common problems with unfeasible parameterisations.

We assume that we have a parameterisation $r_i, a_{ij}$ obtained for biomass abundances $\hat B_i > 0$, survival rates ($r_i=-J_{\min,i}$ for non-basal species $i$), and actual ingestion rates $f_i$ using consumer-resource perspective (see subsections \ref{sec:single}, \ref{sec:feasible}), and the system \eqref{GLV} does not have a feasible equilibrium such that $\dot B_i = 0$, $B_i > 0$ for all $i$. It is irrelevant whether this parameterisation was designed for a steady state (falsely) or for a non-equilibrium state. It is important that for any predator $j$ the feeding rate $f_j$ and coefficients $a_{ji}$ satisfy by definition the relation
$$
f_j = \sum_i \frac{f_j p_{ji}}{\hat B_i} \hat B_i = \sum_i a_{ji} \hat B_i.
$$
Our goal is to obtain a feasible parameterisation at equilibrium ($\dot B_i = 0$ for all $i$) from the same resource-consumer perspective. For that, we introduce a variable $\frac{J_{\min,j}}{f_j}\leq q_j \leq 1$ such that the new feeding rates are $f_j q_j \leq f_j$.
For top-level predators $j$ the equilibrium condition implies $f_j q_j = r_j$, hence $q_j = \frac{r_j}{f_j}$, automatically. For the species $i$ at the next trophic level, we obtain
$$
-r_i + \sum_j \frac{f_j q_j \sigma_{ji} B_j}{e_{ji} \sum_k \sigma_{jk} B_k} = q_i f_i.
$$
We denote $v_j = \frac{q_j  B_j}{\sum_k \sigma_{jk} B_k}$, and solve simultaneously the equations
\begin{equation}\label{eq:opt}
-r_i + \sum_j \frac{f_j v_j \sigma_{ji}}{e_{ji}} = q_i f_i
\end{equation}
for all species $i \in I_1$ that are prey of only top-level predators $j \in J_1$ using the following procedure
\begin{enumerate}
\item For each $i \in I_1$, $j \in J_1$ assume $v_j :=  \hat{v}_i \frac{q_j \hat B_j}{\sum_k \sigma_{jk} \hat B_k}$, $q_i = 1$ and find $\hat{v}_i$ from \eqref{eq:opt}.
\item For each $j \in J_1$, choose $v_j = (\min_{i \in I_1} \hat{v}_i) \frac{q_j \hat B_j}{\sum_k \sigma_{jk} \hat B_k}$.
\item Move $i$ corresponding to minimal $\hat{v}_i$ from the set $I_1$ into the set $\hat I_1$, and move $j$ that are predators of $i$ from $J_1$ to $\hat J_1$.
\item Repeat steps 1-3 until $I_1$ or $J_1$ are empty.
\item For all $i \in I_1 \cup \hat I_1$ find $q_i$ from \eqref{eq:opt}.
\item Choose the set $J_2 = J_1 \cup I_1 \cup \hat I_1$, and the prey $i \in I_2$ with all their predators $j \in J_2$, and if $I_2$ is not empty continue from Step 1 for corresponding sets $I_2, J_2$. For basal species $i$ we explicitly replace $-r_i$ in \eqref{eq:opt} with $J_{\min,i}$.
\end{enumerate}
After that, we obtain all the values $v_j, q_i$, and all equilibrium abundances $B_i > 0$ (where $B_i = \hat B_i$  for the basal species $i$, and all other values are obtained using the formula for $v_j$), and can obtain the coefficients $a_{ij}$ , $r_i > 0$ (for basal species $i$, $r_i < 0$ for non-basal species stay the same) as in the previous subsection.

Therefore, we have demonstrated a way to correct interaction rates of the species in order to obtain feasible parameterisation of GLV. For that, we have to decrease feeding rates at the equilibrium state $B_i> 0$ with respect to the initially chosen ingestion rates $f_i$ (maybe random) at some chosen (non-equilibrium) state $\hat B_i>0$. This suggests that the main source of infeasibility of some common parameterisations is relatively large difference between actual ingestion rates and basal metabolism-related rates for some species at a prospective positive equilibrium state, which cannot be sustained.
We note that this algorithm is not as robust and straightforward as the approach outlined in subsection \ref{sec:feasible}. It tries to maximise minimal values of $B_i$ by maximising minimal values of $v_i$, however it cannot guarantee that all the equilibrium biomass abundances will be near the values $\hat B_i$, whereas in subsection \ref{sec:feasible} usually one can obtain parameterisation for a steady state $B_i = \hat B_i$ by a completely non-random choice of feeding and interaction rates that naturally depend on chosen biomass abundances and basal metabolic rates.
\section{Parameterisation of real food webs}
We consider 8 complex food webs from the data set (\cite{brose2005body}), and in addition two food webs obtained in private communication (\cite{emmerson2004predator,o2009perturbations}), where Lough Hyne food web is an assembled food web (see Table \ref{tablewebs}). For comparison, we also consider a \emph{naive} approach, where the feeding rates $f_j$ are chosen so that basal metabolism-specific rates (survival rates) $J_{min,j}$ are a fixed fraction of feeding rates $f_j$, and the ratio is fixed to 0.4, and birth/death rates $r_i$ are found as $r_i =-\sum_j a_{ij} B_j^*$. (Alternatively, we could parameterise rates $r_i$, and try to solve linear system to find $B_j^*$.)
\begin{table}
  \centering
  \caption{Food webs used in numerical experiments. Columns: index of the food web, place, author/year, number of species,
  number of connections, number of trophic levels}\label{tablewebs}
\begin{tabular}{|l|l|l|l|l|l|}
  \hline
  $N$ & Place & Author, Year & $N_s$ & $N_c$ & $N_l$ \\
  \hline
  1 & UK, Sheffield & Warren & 60 & 166 & 4 \\
  2 & Australia (terrestrial) & Dell & 84 & 599 & 4 \\
  3 & UK, the River Frome, Dorset & Ledger, Edwards, Woodward & 80 & 367 & 2 \\
  4 & Europe, Celtic Sea & Pinnegar, 2003 & 57 & 200 & 8 \\
  5 & Australia, Mulgrave River & Rayner & 62 & 213 & 2 \\
  6 & UK, Silwood Park, Berkshire & Cohen, 2005 & 34 & 56 & 2 \\
  7 & USA, Tuesday Lake & 2004, Jonsson & 73 & 401 & 8 \\
  8 & Antarctica, Eastern Weddell Sea & Jacob, Brey, Mintenbeck & 460 & 1899 & 11 \\
  9 & UK, Ythan Estuary & Emmerson, Raffaelli, 2004 & 87 & 419 & 10 \\
  10 & Ireland, Lough Hyne & O'Gorman,Emmerson, 2009 & 148 & 1295 & 11 \\
  \hline
\end{tabular}
\end{table} 

We see in Table \ref{tableparam} that using the naive approach one can obtain feasible parameterisations of the GLV model for 6 food webs out of 10, where the number of species is less than 80 and the number of connections is less than 410. We see that only  for 4 food webs feasible equilibria can be locally stable  (without non-basal competition), and small intraspecific competition for non-basal species can make these steady states locally stable. We note that maximal eigenvalues that are positive can lead either to the loss of persistence of the system (where some species die out), or to stable dynamics near the unstable equilibrium. In particular,  if the maximal eigenvalues are very close to 0 (up to numerical precision), it can be a strong indication that the system is persistent for all practical situations, because even if some species will eventually die out, it will happen in unrealistically long time. On the contrary, large positive eigenvalues usually lead to the loss of the permanence in the observable time interval.

\begin{table}
  \centering
  \caption{Averaged results of parameterisation of the food webs using naive approach in 1000 runs. The columns: food web number, percent of feasible, maximal number of infeasible $r_i > 0$ for non-basal species, percent of stable for non-basal species without competition, order of maximal eigenvalue, order of maximal eigenvalue for non-basal species with competition $B_i^*/K_i=0.001$ ($K_i$ is the carrying capacity).}\label{tableparam}
\begin{tabular}{|l|l|l|l|l|l|}
  \hline
  $N$ & Feasible & $N(r_i > 0)$ &  Stable & $\lambda_0$ & $\lambda_1$ \\
  \hline
  1 & 99.39\% & 1 & 100\% & $-10^{-9}$ & $-10^{-7}$ \\
  2 & 0\% & 5   & N/A  & N/A & N/A \\
  3 & 100\% & 0 & 0\%&$10^{-20}$ & $-10^{-14}$ \\
  4 & 30.15\% & 1 & 100\%&$-10^{-11}$& $-10^{-8}$ \\
  5 & 100\% & 0 & 6.65\% &  $10^{-22}$ & $-10^{-13}$  \\
  6 & 100\% & 0 & 24.1\%&  $10^{-20}$ & $-10^{-7}$ \\
  7 & 99.53\%& 1 & 0\% & $10^{-10}$ & $-10^{-8}$  \\
  8 & 0\% & 13 & N/A & N/A&N/A \\
  9 & 0\% & 18 & N/A & N/A & N/A \\
  10 & 0\%& 6 & N/A  & N/A & N/A \\
  \hline
  \hline
\end{tabular}
\end{table}

We use the approach to obtain feasible parameterisations (see Table \ref{tableparam2}), and see that the local stability properties of GLV model are similar to the ones obtained by the naive approach. We note that the proportion between the actual $f_i$ and basal $J_{\min,i}$ metabolic rates of non-basal species stays between 1 and 14, and only for the food web 2 it reaches 35. We see that the slope of the allometric relationship $f_i \sim s_i^\beta$ stays close to $\beta \approx -0.25$ similarly to the initially chosen slope for metabolism-related (survival) rates (for all food webs the mean is $\beta \approx -0.23$). We check numerically if the food webs 7, 9, 10 are persistent despite the maximal eigenvalue significantly above 0, and we observe periodic or irregular dynamics with no apparent extinctions. Furthermore, we see that the introduction of weak intraspecific competition $c_{ii}=10^{-4}$ for non-basal species stabilizes the positive steady state. Finally, we check parameterisations of the models with Holling type II and type III functional responses without non-basal competition. We see that most of the positive steady states for most of the food webs are strongly unstable for Holling type II functional response so that some species eventually die out, where only for the first food web the steady state is stable and for the 5th and 6th there are no extinctions for long simulation times. This result agrees with the previous observations (\cite{kalinkat2013body}) and could be explained by the paradox of enrichment and its destabilizing effect on the communities. For Holling type III functional response, however, the resulting food webs seem to be more stable than for Holling type I functional response (GLV model). 

\begin{table}
  \centering
  \caption{Averaged results of parameterisation of the food webs using approach described in section \ref{sec:feasible} in 1000 runs. The columns: food web number, percent of stable for non-basal species without competition, order of maximal eigenvalue $\lambda_0$, order of maximal eigenvalue $\lambda_1$ for non-basal species with competition $c_{ii}=0.0001$, order of maximal eigenvalues $\lambda_{II}$ and $\lambda_{III}$ without non-basal competition for Holling type II and type III functional responses, correspondingly (for a single run), and the constant $\beta$ for allometric relationship for the feeding rate $f_i \sim s_i^\beta$ with the confidence interval $[\beta_l, \beta_r]$ for 1000 runs.}
\label{tableparam2}
\begin{tabular}{|l|l|l|lll||lll|lll}
  \hline
  $N$ & Stable & $\lambda_0$ & $\lambda_1$ & $\lambda_{II}$ & $\lambda_{III}$ & $\beta$ & $\beta_l$ & $\beta_r$ \\
  \hline
  1 & 100\% & $-10^{-4}$ & $-10^{-3}$ & $-10^{-4}$ & $-10^{-4}$ &-0.288 & -0.29  &-0.286\\
  2 & 0\% & $10^{-15}$  & $-10^{-3}$  & 10  & $-10^{-16}$ & -0.2068 &  -0.2073 & -0.2062 \\
  3 & 0\%&$10^{-16}$ & $-10^{-3}$ & $10^{-2}$ & $10^{-17}$&-0.236  & -0.237  & -0.235 \\
  4 & 100\%&$-10^{-5}$& $-10^{-3}$ & $10^{-2}$& $-10^{-5}$ & -0.381 & -0.382  &-0.379\\
  5 & 7.1\% &  $10^{-15}$ & $-10^{-3}$ & $10^{-17}$& $10^{-16}$ & -0.2497 &  -0.2499 &  -0.2495  \\
  6 & 23.5\%&  $10^{-16}$ & $-10^{-3}$ & $10^{-16}$& $10^{-16}$ & -0.054 & -0.064 & -0.043\\
  7 &  0\% & $10^{-4}$ & $-10^{-3}$ & $10^{-3}$& $-10^{-3}$ & -0.2607 &  -0.2613  &-0.2602 \\
  8 &  0\% & $10^{-14}$ & $-10^{-3}$ & $10^{-1}$ & $10^{-14}$  & -0.2126 &  -0.2129 &  -0.2123\\
  9 &  0\% & $10^{-4}$ & $-10^{-3}$ & 10 & $10^{-17}$  & -0.1591 &  -0.1597 &  -0.1585\\
  10 &  0\% & $10^{-4}$ & $-10^{-3}$ & $10^{-1}$ & $10^{-16}$ & -0.2455 &  -0.2462 &  -0.2445 \\
  \hline
\end{tabular}
\end{table}

\section{Conlusion}

We have presented an approach to obtain feasible parameterisations for general predator-prey models of real food webs and tested this approach on 10 food webs for the predator-prey systems with Holling type I, type II and type III responses. Using numerical simulations and stability analysis of the positive steady state of the parameterised models we could obtain some preliminary evidence that such models can be persistent for Holling type I and type III functional responses, and that weak intraspecific competition for non-basal species can make positive steady state of the generalised Lotka-Volterra model locally stable. Similarly to the previous works, we observed that the models of real food webs with Holling type II functional response are not persistent.

Finally, we note that though the generalised Lotka-Volterra model, which lies in the focus of this paper, is the simplest dynamical model of food web evolution, and probably the correct answer for the realistic persistence of the food webs lies in much more involved modelling (\cite{kalinkat2013body,allhoff2014evolutionary,schneider2016animal}), it allows for steady-state analysis and can provide us with general insights that are much harder to obtain using more realistic models (\cite{berg2015ecological,jonsson2015reliability,jonsson2015context}). We hope that the presented approach will facilitate the research of the dynamics of the real food webs.

\section{Acknowledgements}
I would like to thank Mark Emmerson and IRCSET for the opportunity to dive deep into the world of complex population models, and for long discussions that led eventually to the appearance of this paper. I would like also to thank Sofia Berg, Tomas Jonsson, and Catherine Palmer for the opportunity to participate in their projects and for giving me motivation to finish this work. Finally, I acknowledge the financial support of my current research activities from SFB 787 of the DFG.

\bibliographystyle{apalike}       

\bibliography{foodweb}

\end{document}